\documentclass[notitlepage,superscriptaddress,showpacs,nobalancelastpage,twocolumn,aps,longbibliography,prl]{revtex4-2}
\usepackage[english]{babel}
\RequirePackage[T1]{fontenc}

\usepackage{siunitx} 
\usepackage[italicdiff]{physics}
\usepackage{amsfonts}
\usepackage{lipsum}
\usepackage{multirow}
\usepackage{amsmath}
\usepackage{amssymb}
\usepackage{graphicx, epstopdf}
\usepackage{mathtools}
\usepackage{wasysym}
\usepackage{appendix}
\usepackage{xspace}
\usepackage{array}
\usepackage[dvipsnames]{xcolor}
\usepackage{my_symbols}
\usepackage{CJKutf8}
\usepackage{bm}
\usepackage{verbatim}
\usepackage{tikz}
\usepackage{comment}
\usepackage{marginnote}
\usepackage[textwidth=1.5cm]{todonotes}
\usepackage[normalem]{ulem}
\usepackage{soul}
\usepackage[commandnameprefix=ifneeded]{changes}
\usepackage{mathrsfs}
\usepackage[hidelinks]{hyperref}

\sethlcolor{green}

{\par}

\newcommand{\ii}{\mathrm i}
\renewcommand{\dd}{\mathrm d}
\newcommand{\cmmnt}[1]{}
\newcommand{\hcj}{\text{h.c.}}

\excludecomment{comment}  

\begin{document}
\begin{CJK*}{UTF8}{gbsn} 
\title{Quantum Entanglement and Teleportation of Magnons in Coupled Spin Chains}
\author{Zian Xia}
\author{Ruoban Ma}
\author{Chang Shu}
\thanks{Current Address: Department of Physics, The University of Michigan, USA}
\affiliation{State Key Laboratory of Surface Physics and Department of Physics, Fudan University, Shanghai 200433, China}
\author{Huaiyang Yuan}
\affiliation{Institute for Advanced Study in Physics, Zhejiang University, 310027 Hangzhou, China}
\author{Jiang Xiao (萧江)}
\email[Corresponding author:~]{xiaojiang@fudan.edu.cn}
\affiliation{State Key Laboratory of Surface Physics and Department of Physics, Fudan University, Shanghai 200433, China}
\affiliation{Institute for Nanoelectronics Devices and Quantum Computing, Fudan University, Shanghai 200433, China}
\affiliation{Shanghai Research Center for Quantum Sciences, Shanghai 201315, China}
\affiliation{Hefei National Laboratory, Hefei 230088, China}

\begin{abstract} 

    This study explores how entanglement and quantum teleportation of magnons can be achieved in coupled spin chain systems. By utilizing different magnetic configurations, we show that parallel spin chains function like magnonic beam splitters, whereas anti-parallel chains produce two-magnon squeezing and strong entanglement. Combining these components, we design magnonic circuits capable of continuous-variable quantum entanglement and teleportation, supported by quantum Langevin simulations. 

\end{abstract}

\maketitle
\end{CJK*}

\emph{Introduction --} 
Quantum information and computation exploit inherently quantum features \cite{Nielsen_Chuang_2010}, notably superposition and entanglement, to achieve information processing tasks unattainable by classical means. In recent decade, experimental advancements in diverse platforms, ranging from superconducting circuits \cite{Josephson_Persistent-Current_Qubit_1999,Superconducting_Flux_Qubit_2003,Superconducting_circuits_fault_tolerance_2014,2013_Coherent_Josephson_qubit_scalable_circuits} and trapped ions \cite{Quantum_Computations_with_Cold_Trapped_Ions_1995,ion_trap_quantum_information_processor_1997,The_physical_implementation_quantum_computation_2000} to photonic architectures \cite{Cavity-QED_universal_quantum_gate_1995,Cavity_QED_Model_1995,Measurement_Phase_Shifts_Quantum_Logic_1995,Quantum_computation_linear_optics_2001} and nitrogen-vacancy in diamonds \cite{Processing_quantum_information_diamond_2006,Quantum_Registe_Diamond_2007,Quantum_computing_with_defects_2010,Scalable_Quantum_Information_Processing_Diamond_2014}, have substantially improved coherence, scalability, and operational control. 

Magnonics is a rapidly developing field focused on the study of spin waves -- collective excitations of spins in magnetic materials -- describable as both classical waves and quantized quasi-particles known as magnons \cite{demokritovMagnonicsFundamentalsApplications2012,zarerameshtiCavityMagnonics2022}. Magnonic systems stand out due to their low intrinsic energy dissipation and high tunability. More importantly, its capability to interact with other quasiparticles, such as microwave photons, phonons, and superconducting qubits, makes magnon attractive for advancing quantum information technologies \cite{lachance-quirionHybridQuantumSystems2019,Entanglement_anisotropic_magnets_2020,2021_Magnon_magnon_entanglement,yuanQuantumMagnonicsWhen2022,zarerameshtiCavityMagnonics2022}. 
In the past decade, the strong coherent coupling has been demonstrated between magnons in yttrium iron garnet (YIG) spheres and cavity photons \cite{hueblHighCooperativityCoupled2013,goryachevHighCooperativityCavityQED2014,tabuchiHybridizingFerromagneticMagnons2014,baiSpinPumpingElectrodynamically2015,zhangObservationExceptionalPoint2017,Cavity_magnonics_Review}, as well as indirect coupling between magnons and superconducting qubits mediated by cavity photons \cite{tabuchiCoherentCouplingFerromagnetic2015,lachance-quirionResolvingQuantaCollective2017}. 
More recently, the deterministic creation \cite{Single_magnon_detection_2020} and control \cite{xuQuantumControlSingle2023} of quantum magnonic states have been realized in cavity magnonic systems. The tomography techniques for magnon Wigner distribution have also been achieved for both classical \cite{Tomography_Parametric_Transition_Magnets_2025,Persistent_magnetic_coherence_2024} and quantum states \cite{xuQuantumControlSingle2023}. 
These advances highlight the promising role of magnons in enabling new applications in quantum information processing and quantum computing.

Current research using magnetic system
for quantum information processing mainly focus on the hybrid systems involving magnons (mostly the Kittel mode in YIG spheres) and microwave photons \cite{Magnon_Photon_Phonon_Entanglement_2018,2021_Magnon_magnon_entanglement,Enhancement_magnon_entanglement_cavity_2020}. 
Other attempts include building qubit based on the chirality of magnetic domain wall \cite{Domain_wall_qubits_2023} or the helicity of a skyrmion \cite{Skyrmion_Helicity_Qubits_2023}, both explore the potential of non-collinear magnetic texture \cite{yuMagneticTextureBased2021}.
In contrast, approaches relying solely on magnonic systems, especially those based on propagating magnons, remain relatively unexplored. 
This paper investigates theoretically the potential of propagating magnons in quantum information tasks. Specifically, we examine entanglement generation between two magnonic excitations and the teleportation of magnon state, utilizing a quasi-one-dimensional magnonic system formed by coupled spin chains or magnetic nanowires. In this approach, two transverse components of bosonic magnonic excitations serve as carriers of continuous quantum information \cite{Quantum_information_with_continuous_variables_Review_2005}, analogous to the roles of quadrature components of photons in quantum optics or optomechanics \cite{Quantum_computation_linear_optics_2001,Quantum_computation_with_optical_coherent_states_2003,Optomechanical_Entanglement_Mirror_Cavity_2007}. Most interestingly, we find that, depending on the relative alignment of the equilibrium magnetization in the two chains, the inter-chain exchange interaction can realize the functionality of a beam splitter or of a two-mode squeezer. This magnetic reconfigurability may simplify the construction of quantum magnonic circuits.

\emph{Model --}
We investigate the propagation of magnon wave packets along a quasi-one-dimensional system consisting of two magnetic chains. In certain regions, the proximity of the two chains allows for an exchange-like interaction across the two chains. The Hamiltonian includes the contribution from two independent chains \(\hH_0\) and the coupling \(\hH_c\) between two chains:  
\begin{equation}  
    \hH = \hH_0 + \hH_c.  
\end{equation}  
$\hat{H}_0$ includes the easy-axis anisotropy (in $\vu{z}$),
and the ferromagnetic intra-chain exchange coupling ($J < 0$), 
\begin{equation*}
    \hH_0 = \sum_{\alpha = A, B}   
    \int \dd{x}
    \qty{-\frac{K}{2} \qty[\mb_\alpha(x) \cdot \hbz]^2
    + \frac{A}{2} \qty[  \pdv{\mb_\alpha (x)}{x}]^2},  
\end{equation*}
where $\mb_\alpha(x)$ denotes the normalized magnetic moment at position $x$ on chain-$\alpha$, $K$ and $A$ are the uniaxial anisotropy (along $\hbz$) and the exchange constant. 
The inter-chain coupling Hamiltonian is expressed as:  
\begin{equation*}  
    \hH_c = J' \int_{\mathcal{C}} \dd{x}~\mb_A(x) \cdot \mb_B(x),  
\end{equation*}  
where \(J'\) is the inter-chain coupling strength and \(\mathcal{C}\) denotes the region where the two chains are coupled.

In order to carry out numerical simulations, we discretize the Hamiltonian into $N$ segments in the $x$ direction along the chain of length $L$, and let $\mb_\alpha^i = \mb_\alpha(x_i)$, where $L/N$ is the discretization length within which all spins are effectively regarded as a single moment.
Employing the Holstein-Primakoff (H.P.)
transformation \cite{Holstein_Primakoff_1940} along the equilibrium magnetization direction (denoted as $\mb_{\alpha,0}^{i}$) of each chain facilitates the derivation of the effective Hamiltonian for low energy magnon excitations \cite{stancilSpinWaves2009}:
\begin{equation}
    \hH_0^{\text{eff}} = \sum_{\alpha,i} K \bar{m}_\alpha^i m_\alpha^i  
    + J \qty(\bar{m}_\alpha^i m_\alpha^{i+1} + m_\alpha^i \bar{m}_\alpha^{i+1}),
\end{equation}
where $m_\alpha^i =( m_\alpha^{i,X} + i m_\alpha^{i,Y} )/\sqrt2$ and $\bar{m} \equiv m^\dagger$ are the local magnon annihilation (creation) operator at site-$i$ on chain-$\alpha$, and $J = - A (N/L)^2$
is the renormalized intra-chain Heisenberg exchange constant for the discretized model. Here $Z$ is chosen to be aligned with the local equilibrium magnetization direction ($\mb_{\alpha,0}^{i}$), which is not necessarily to be the same as the easy-axis $\hat{z}$ because of the possible existence of a magnetic domain wall in the chain.

\begin{figure}[t]
    \includegraphics[width=\columnwidth]{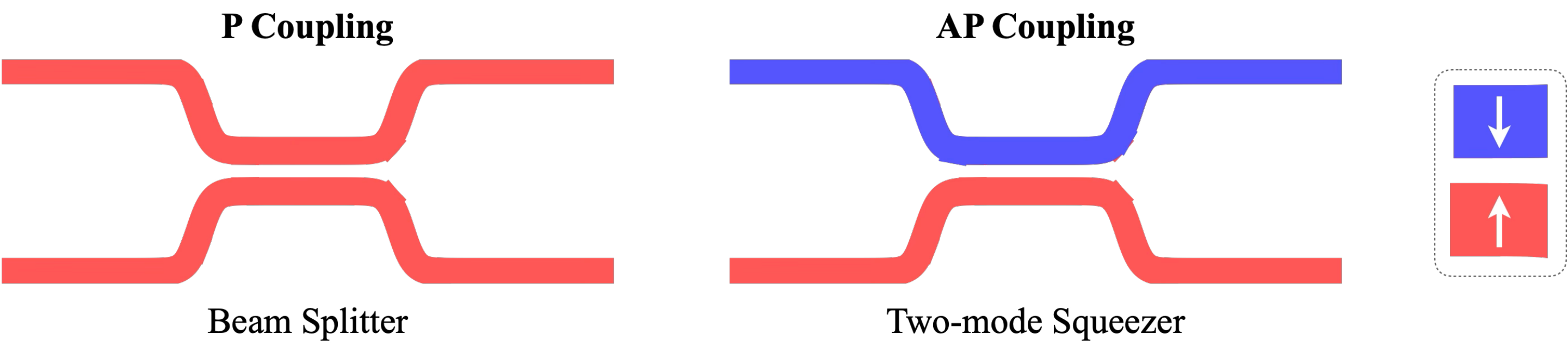}
    \caption{The realization of beam splitter and two-mode squeezer for magnons using parallel and anti-parallel configurations.}
    \label{fig:model} 
\end{figure}

The inter-chain coupling $\hat{H}_c$ exhibits distinct behaviors depending on the relative alignment of their equilibrium magnetization. When the equilibrium magnetization in the two chains within the coupling region ($\mathcal C$) are parallel (P, $\mb_{A,0}^i = \mb_{B,0}^i, i\in {\mathcal C}$, see \Figure{fig:model}(a)), the coupling Hamiltonian 
\begin{equation}
    \label{eqn:HP}
    \hH_c^{\text{P}} 
    = J' \sum_{i\in \mathcal{C}} \bar{m}_A^i m_B^i + m_A^i \bar{m}_B^i,
\end{equation}
takes the form of a beam-splitter interaction. However, when the equilibrium magnetization are anti-parallel (AP, $\mb_{A,0}^i = -\mb_{B,0}^i, i\in {\mathcal C}$, see \Figure{fig:model}(b)), the coupling Hamiltonian is a parametric down-conversion-like or two-mode squeezing interaction \cite{Antiferromagnetic_magnons_squeezed_Fock_states_2019}
\begin{equation}
    \label{eqn:HAP}
    \hH_c^{\text{AP}} 
    = J' \sum_{i\in \mathcal{C}} m_A^i m_B^i + \bar{m}_A^i \bar{m}_B^i.
\end{equation}
Comparing with other bosonic systems, where achieving two-mode squeezing usually requires complex nonlinear processes \cite{Number_phase_minimum_uncertainty_states_1987}, the two-mode squeezing interaction in magnonic systems, however, can be realized simply by arranging the equilibrium magnetic configuration into an antiparallel alignment \cite{Enhancement_magnon_entanglement_cavity_2020,Antiferromagnetic_magnons_squeezed_Fock_states_2019}. It is important to note that the emergence of the two-mode squeezing interaction here only relies on whether the equilibrium magnetizations are aligned parallel or antiparallel, and is not affected by whether the inter-chain coupling is ferromagnetic ($J' < 0$) or antiferromagnetic ($J' > 0$).

\begin{figure}[t]
    \centering
    \includegraphics[width=\columnwidth]{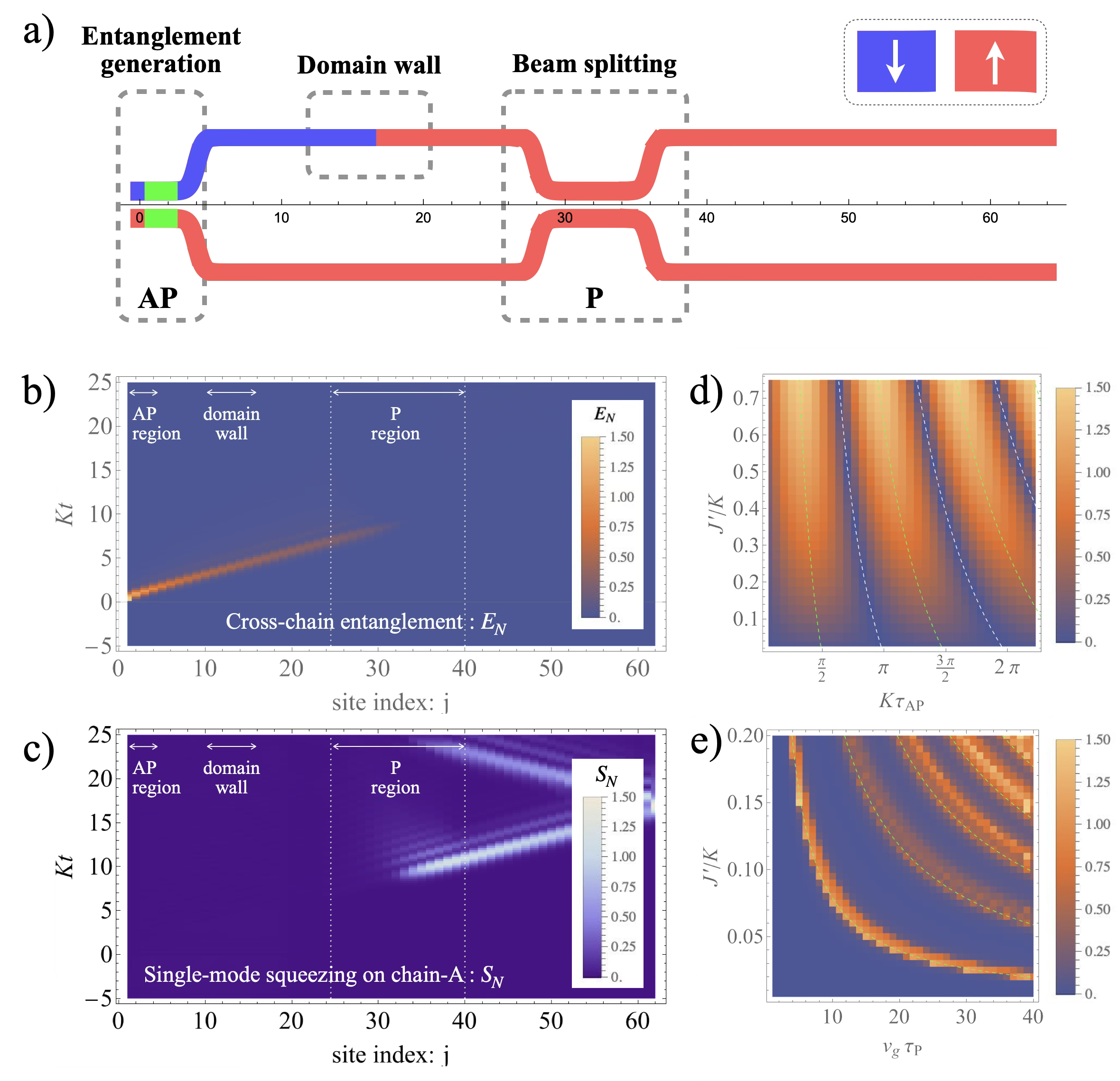}
    \caption{
        (a) The quantum circuitry for entanglement generation (in AP region) and conversion to single-mode squeezing (in P region).
        (b) The temporal-spatial propagation of cross-chain entanglement.
        (c) The temporal-spatial propagation of the single-mode squeezing (in chain-$A$).
        (d) The magnitude of the entanglement evaluated around site-10 as function of coupling strength $J'$ and duration $\tau_{\text{AP}}$ in AP.
        (e) The residue entanglement evaluated around site-45 as a function of $J'$ and $\tau_{\text{P}}$.
        }
    \label{fig:doublechain}
\end{figure}

In order to realize both P and AP type coupling in the same double-chain system (an example is shown in \Figure{fig:doublechain}(a)), a magnetic domain wall can be introduced in one of the chains (say chain-B). The existence of a Walker type domain wall on chain-B 
introduces an additional term to the Hamiltonian (see Appendix A):
\[
\begin{aligned}
    \hH^{\text{DW}} &= \sum_{i} -\frac{K}{2} \sin^2\theta_i (3\bar{m}_B^i m_B^i + \qty(\bar{m}_B^i \bar{m}_B^i + \hcj) ) \\
    & + J \sin^2\frac{\Delta\theta_i}{2}\qty[ 2 \bar{m}_B^i m_B^i + \qty((m_B^i  + \bar{m}_B^i) m_B^{i+1} + \hcj) ] 
\end{aligned}
\]
where $\theta_i$ is the angle between the equilibrium magnetizations of site-$ i $ and the anisotropy axis, and $\Delta\theta_i = \theta_{i+1} - \theta_i$ is the relative angle between the equilibrium magnetizations on the neighboring $i$-th and the $i+1$-th sites on chain B.

\emph{Quantum Langevin Equation --}
Since the full Hamiltonian is quadratic in $m$ and $\bar{m}$, the corresponding Heisenberg equation of motion for these operators are linear:
\begin{equation}
    \label{eqn:EOM}
    \dv{t}\hat{M} 
    = \ii \comm*{\hat H}{\hat{M}} 
    = \ii D \hat{M},
\end{equation}
where $\hat{M}$ is the collection of all magnon creation/annihilation operators along the two chains with $\hat{M}_{\alpha i +} = \bar m_{\alpha}^i$ and $\hat{M}_{\alpha i -} = m_{\alpha}^i$, and $s = \pm$ denoting the creation and annihilation operator. 
Here $D$ is a $4N\times 4N$ drift matrix whose elements are given by (see Appendix A)
\begin{equation}
    \label{eqn:driftmtx}
    D_{\alpha is}^{\alpha'i's'} 
    = s K
    \Delta_{\alpha i s}^{\alpha'i' s'}
    + s J \Delta_{\alpha i_\pm s}^{\alpha' i' s'}
    + s J' \begin{cases}
    \Delta_{\overline{\alpha} i s}^{\alpha' i' s'} &\mbox{(P)}, \\
    \Delta_{\overline{\alpha} i \overline{s}}^{\alpha' i' s'} &\mbox{(AP)}.
    \end{cases} 
\end{equation}
Here $\Delta_{\alpha i s}^{\alpha'i' s'} = 1$ when $\alpha, i, s = \alpha' , i' , s'$ and 0 otherwise, $i_\pm = i\pm 1, \bar{s} = -s$, and $\bar\alpha$ represents the opposite chain of $\alpha$ ($\overline{A} = B, \overline{B} = A$). 
When a domain wall is present, an addition term $D^{\text{DW}}$ corresponding to $\hat{H}^{\text{DW}}$ should be included (see Appendix A).

Because the system is Gaussian with purely quadratic Hamiltonian, the Heisenberg equation \Eq{eqn:EOM} for the quantum operator $\hat{M}$ can be reformulated as a stochastic equation for a classical random variable $M$, known as the quantum Langevin equation (QLE) \cite{Quantum_Langevin_equation_1988,Approaches_to_quantum_dissipative_systems}:
\begin{equation}
    \label{eqn:EOMX}
  \dv{t} M(t) = D M(t) + h(t),
\end{equation} 
where $M_{\alpha i s}(t)$ is a time dependent random variable subject to a Gaussian distribution, and $h_{\alpha is}(t)$ is the stochastic noise to emulate quantum fluctuation with correlation 
$\langle h_{\alpha i s}(t)h_{\alpha' i' s'}(t') \rangle = \Delta_{\alpha i \overline{s}}^{\alpha' i' s'}\delta(t-t')$ \cite{Quantum_Langevin_equation_1988}.  
The equivalence between \Eq{eqn:EOM} and \Eq{eqn:EOMX} is because the ensemble average over the random variables $\ev*{M}_{\text{random}}$ reproduces exactly the expectation value of the quantum operator $\ev*{\hat{M}}_{\text{quantum}}$ for Gaussian states \cite{Quantum_information_with_continuous_variables_Review_2005}, therefore it is sufficient to solve the QLE for $M$ to obtain the quantum information carried in quantum operator $\hat{M}$ \cite{Quantum_Langevin_equation_1988}. For instance, the first-order moment of the QLE, $\ev*{\dot{M}} = D\ev*{M}$, agrees with the deterministic dynamics described by the linearized Landau-Lifshitz-Gilbert (LLG) equation.
The influence of finite temperature and dissipation are not included in \Eq{eqn:EOMX}, but are considered as an additional Gaussian noise and discussed in Appendix B.

For Gaussian state magnons, all quantum properties including squeezing and entanglement, are encoded in the second order moment of $M$ described by the covariant matrix \cite{Gaussian_quantum_information_2012}: 
\[ V_{\alpha i s}^{\alpha'i's'} 
= \ev{M_{\alpha i s} M_{\alpha'i's'}} 
- \ev{M_{\alpha i s}} \ev{M_{\alpha'i's'}}, \] 
whose equation of motion follows from \Eq{eqn:EOMX} \cite{Enhancement_magnon_entanglement_cavity_2020,EPR_paradox_in_pulsed_optomechanics_2013}:
\begin{equation}
  \label{eqn:dVdt}
  \dv{t}V(t) = DV(t) + V(t)D^T.
\end{equation}
With the full covariant matrix $V(t)$ solved from \Eq{eqn:dVdt}, the entanglement between the magnon excitation can be extracted from $V$ by using the logarithmic negativity \cite{Gaussian_quantum_information_2012}: 
\begin{equation}
    \label{eqn:EN}
  E_N[V] = \max (0,-\ln 2\nu_-), 
\end{equation} 
where $\nu_-$ is the minimum symplectic eigenvalue of the partially transposed covariant matrix $\tilde{V}$. The partial transpose is effectively 
swapping $M_{Ai\pm} \leftrightarrow M_{Ai\mp}$ (or $m_A^i \leftrightarrow \bar{m}_A^i$) on chain $A$ only while variables on chain $B$ are unchanged, so that 
$
\tilde{V}_{Aks}^{Ak's'} = V_{Ak\overline s}^{Ak'\overline{s'}},
\tilde{V}_{Aks}^{B'k's'} = V_{Ak\overline s}^{Bk's'},
\tilde{V}_{Bks}^{Bk's'} = V_{Bks}^{Bk's'}.$
Because the partial transpose is applied only on one of the chains (chain-$A$ here), only the inter-chain entanglement contributes to $E_N$. The intra-chain entanglement, if exists, is not included in $E_N$.
Furthermore, we may also measure the squeezing within chain-$A$ by the squeezing parameter defined as \cite{Magnon_squeezing_niche_2020,Squeezing_versus_photon_number_fluctuations_1987}:
\begin{equation}
  \label{eqn:xi}
  S_N[V_A] = \max (0, -\ln \lambda_-^A),
\end{equation}
where $V_A$ is the submatrix of $V$ restricted on chain-$A$, and $\lambda_-^A$ is the smallest eigenvalue of $V_A$ (similar definition for $V_B$). 

To investigate the temporal evolution of the spatial distribution of entanglement across the double-chain system, we introduce a projection operator $P_\mathcal{D}$ that projects the full covariance matrix $V$ onto a specific subspace $\mathcal{D}$, denoted as $V_\mathcal{D} \equiv P_\mathcal{D} V$, representing a submatrix of $V$ with restricted indices $\alpha i s, \alpha' i' s' \in \mathcal{D}$. 
For instance, in \Eq{eqn:xi}, $V_A = P_A V$ is the single-chain submatrix with $A = \{\alpha i s~|~\alpha = A \}$. 
To specifically analyze the spatial distribution of inter-chain entanglement, we define the submatrix $V_j \equiv P_{\mathcal{D}_j} V$, which isolates a window of width $w$ centered at site $j$ across both chains with $\mathcal{D}_j = \{\alpha i s~|~\abs{i-j} \le w/2 \}$.
If we further restrict this window to be on one particular chain (such as chian-$A$), we define $V^A_j \equiv P_A V_j$. 
Consequently, both the inter-chain entanglement and single-chain squeezing within this moving window can be quantified using the derived submatrices as $E_N[V_j]$ and $S_N[V_j^A]$ in \Eqs{eqn:EN}{eqn:xi}, respectively.

\emph{Quantum Circuit for Entanglement Verification --}
A magnonic double-chain architecture, as shown in \Figure{fig:doublechain}(a), leverages engineered coupling regions to both generate and diagnose quantum correlations between two propagating magnonic wave packets. The device incorporates an AP coupling region followed downstream by a P coupling region, with the transition between them controlled by a 180$^\circ$ magnetic domain wall in the upper chain. We assume that the coupling strength within the AP region can be turned on and off on demand. By activating the AP coupling for a short period of time, the initially trivial vacuum state in the AP region of the double chain is converted into a two-mode squeezed vacuum across the double chain by $\hat{H}_c^\text{AP}$, thereby establishing entanglement across the double-chain. The resulting entangled vacuum state then propagate as two magnonic wave packets independently along each chain to the right, preserving their quantum correlations even as the upper packet traverses the domain wall. The subsequent P-coupled section acts as a beam splitter. Given the beam-splitting phase to be $\pi/4$, the two-mode entanglement across the double-chain is transformed into two spatially separated modes that exhibit single-mode squeezing. Because the single-mode squeezing is localized within each chain, it can be measured via local quadrature-sensitive probes, enabling practical verification of the generated entanglement without requiring nonlocal joint detection.

We simulate the magnon generation and propagation in the double chain illustrated above using the QLE in \Eq{eqn:dVdt}.
In the simulation, each chain contains $N = 64$ sites. The AP coupling region covers from site-1 to -3, and the P coupling region covers site-25 to -40, with the length $l_{\text{P}} = 16$. The magnetic domain wall in the upper chain is positioned between site-13 to -20.
The coupling in the AP region is briefly turned on for a short duration $0 < t < \tau_{\text{AP}} $ to create the squeezed vacuum across the double chain.
The value of $\tau_{\text{AP}}$
is chosen such that maximum squeezing is realized.
The inter- and intra-chain coupling strength are $\abs{J}=0.125 K, J'=0.3K$.
Here $\abs{J}=A(N/L)^2<J'$ is required so that the effects of the P/AP coupling act on the wave packet before leaving the P/AP coupling regions. The wave packet has the characteristic length $\sim L/N$, comparing with which, spin waves with shorter wave length and faster group velocity $v_g=2|J|L/N$ are neglected.

\Figure{fig:doublechain}(b) plots the temporal-spatial map of the entanglement $E_N[V_j(t)]$ as function of window-position $j$ and time $t$, which shows that the cross-chain entanglement, generated in the AP region during time $t\in (0,\tau_{\text{AP}})$, travels along the double-chain together with the wave packets and remains confined within the packets, exhibiting relatively weak broadening over time. Interestingly, the presence of a magnetic domain wall around site-13 in the upper chain does not impede or diminish the cross-chain entanglement shared between the wave packet pairs. This behavior is consistent with the fact that the domain wall only leads a phase shift effect within the upper chain, which simply redistributes the entanglement among its magnetic excitation components without reducing the total entanglement across the double chain \cite{Squeezing_irreducible_resource_2005}.

As the wave packet pair passing the P coupling region near site-32, the cross-chain entanglement diminishes (\Figure{fig:doublechain}(b)), while the single-mode squeezing appears instead (\Figure{fig:doublechain}(c)). This change occurs because the P coupling functions as a beam splitter, converting the initial two-mode squeezing or entanglement shared between chains into localized single-mode squeezing. Consequently, quantum information that was once distributed across the double-chain becomes confined within each individual chain. This localization enables the verification of quantum entanglement generated previously using local measurements on either chain-$A$ or chain-$B$.

The magnitude of cross-chain entanglement generated in the AP region is governed by the inter-chain coupling strength $J'$ and the activation duration $\tau_{\text{AP}}$. As shown in \Figure{fig:doublechain}(d), the entanglement (examined at site-10) oscillates as function of $\tau_{\text{AP}}$, reaching its maximum when $\phi_{\text{AP}} = \omega_{\text{AP}} \tau_{\text{AP}} = n\pi + \pi/2$ (green dashed curves) with $\omega_{\text{AP}} = \sqrt{K^2+2KJ'}$. 
The efficiency of beam splitting in the P coupling region also relies on $J'$ and the time $\tau_{\text{P}} = l_{\text{P}} / v_g$ for the wave packet passing through the P region. As shown in \Figure{fig:doublechain}(e), the residual entanglement (examined at site-45) oscillates with $\phi_\text{P} = J'\tau_{\text{P}}$, and optimal beam splitting marked by minimal entanglement occurs around $J'\tau_{\text{P}} \approx n\pi + \pi/4$ (green dashed curves).

\begin{figure}
\includegraphics[width=\columnwidth]{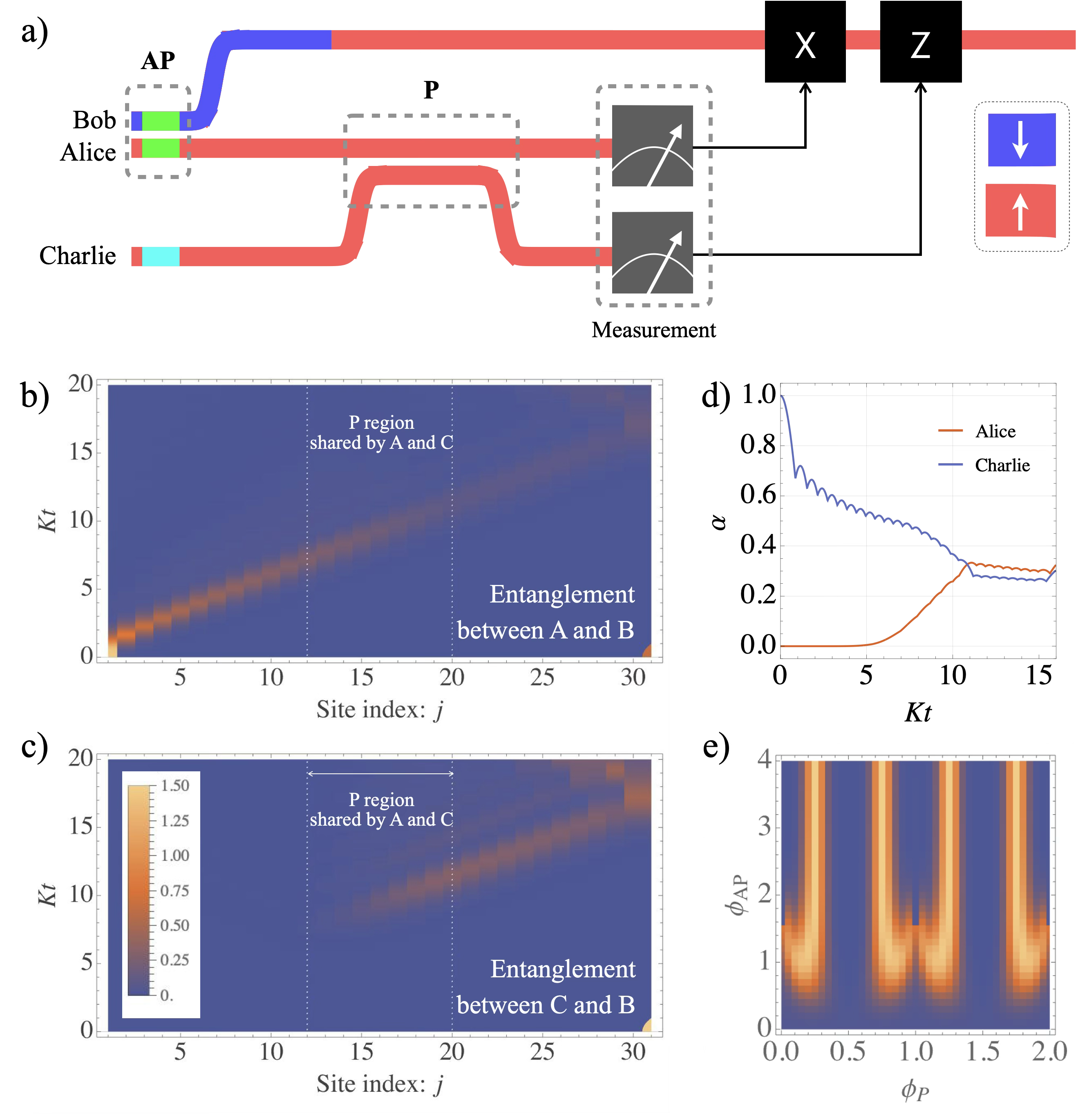}
\caption{ 
    (a) The quantum circuitry for magnon teleportation with three spin chains.  
    (b) The entanglement between chain-A and -B.  
    (c) The entanglement between chain-C and -B.  
    (d) The amplitudes of coherent states on chain-A and -C. 
    (e) The teleportation fidelity as function of the phases realized in the AP and P region. 
    }
\label{fig:tel}
\end{figure}

\emph{Magnon Teleportation --}
We now turn to the realization of a more sophisticated magnonic circuit capable of implementing the continuous variable quantum teleportation protocol. The circuit involves three magnetic chains (A, B, C), designated as Alice, Bob, and Charlie depicted schematically in \Figure{fig:tel}(a). 
Each chain is simulated with 32 sites and $\abs{J} = 0.25K$.
The teleportation process begins with the generation of an entangled magnonic wave packet pair in the AP region shared by Alice and Bob (from site-1 to site-3 with $J'=0.2K$). 
In the meantime, a coherent magnonic wave packet with unit amplitude, as a quantum state intended to be teleported, is excited into Charlie from its left end, by setting $M_{C 1 \pm}(0) = 1/\sqrt{2}$.
A crucial interaction happens when Alice and Charlie meets in a P coupling region (from site-13 to -20 with the length $l_{\text{P}}=8$), where a beam-splitting operation occurs between the wave packets from Alice and Charlie. This mechanism enables partial transfer of the original entanglement shared by the AB pair to the BC pair. 
This partial transfer is evidenced by spatio-temporal maps of entanglement presented in \Figure{fig:tel}(b) and (c), which show that, once entering the P coupling region, the entanglement between Alice and Bob (\Figure{fig:tel}(b)) decreases (but not to zero), while new entanglement between Bob and Charlie (\Figure{fig:tel}(c)) builds up. This demonstrates the successful partial transfer of quantum correlations required for teleportation. Simultaneously, as depicted in \Figure{fig:tel}(d), the quantum information initially encoded in Charlie is also partially relayed to Alice, indicated by an emerging amplitude in Alice and a corresponding decrease in wave packet amplitude in Charlie. 
After the beam splitting between chain-A and -C, independent homodyne detection \cite{Entangled_States_Homodyne_Detector_2009,Nonlinear_cavity_magnonics_quantum_information_2020} measurements are performed on the different components on the two chains ($X$ component on chain-A and $Y$ component on chain-C for instance). The results of these homodyne measurements are then used to apply a phase-space displacement to the state in chain-B, employing X and Z gates, which could be implemented using magnetic field pulses. This displacement operation adjusts the state in chain-B so that it recovers the original coherent state with finite amplitude that was initially input into chain-C. This completes the quantum teleportation process transferring the coherent state from chain-C to chain-B.

\Figure{fig:tel}(e) shows the teleportation fidelity as a function of the phase accumulated in the P region, $\phi_{\text P} = J' \tau_{\text{P}}$, and the phase accumulated in the AP region, $\phi_{\text{AP}} = \omega_{\text{AP}} \tau_{\text{AP}}$. 
Fidelity is evaluated using the overlap between the Wigner functions of the teleported state at Bob and the initial state at Charlie \cite{Optimal_fidelity_of_teleportation_of_coherent_states_2008,Quantum_information_with_continuous_variables_Review_2005}. These Wigner functions are constructed from the first and second moments derived via QLE in \Eq{eqn:dVdt} (see Appendix C). Ideal teleportation requires i) strong entanglement between Alice and Bob, achieved by maximizing $\phi_{\text{AP}}$, and ii) a 50:50 beam-splitter interaction, which occurs when $\phi_{\text P} = \frac\pi4 + n\frac\pi2$. The results in \Figure{fig:tel}(e) demonstrate that these conditions indeed yield the highest teleportation fidelity.

\emph{Discussion and conclusions --} 
The parameters used in our discretized model are based on the material parameters of yttrium iron garnet (YIG): gyromagnetic ratio $\tilde\gamma = \SI{2.21e5} {m.s^{-1}.A^{-1}}$, exchange coupling strength $\tilde A = \SI{3.28e-11} {A.m}$ \cite{YIG_spinwave_2006}, and anisotropy $\tilde K= \SI{3.88e4}{A.m^{-1}}$ \cite{Spin_Transfer_Torque_and_Domain_Wall_Propagation_2011}. 
These parameters are translated into the effective coupling $J = \tilde \gamma \tilde A (N/L)^2$, effective anisotropy $K=\tilde\gamma \tilde K$, yielding a value of $J/K = 0.125$ for $N=64$ and a system size of $L\approx \SI{5}{\mu m}$. The entanglement verification or teleportation process concludes within approximately $50/K \simeq \SI{5}{ns}$. 
The intra-chain coupling ratios range is $J'/K\approx 0.1 \sim 0.5$ in our simulation, which is achievable in synthetic antiferromagnets like CoFeB/Ru/CoFeB layers \cite{Exchange_energies_synthetic_antiferromagnets_2023}. 
Furthermore, the system demonstrates robustness against thermal noise. At temperatures around $\SI{65}{mK}$, corresponding to a single thermal magnon excitation in YIG, entanglement remains stable (see Appendix B), a condition readily attainable in current experimental setups \cite{Single_magnon_detection_2020,xuQuantumControlSingle2023}.

In summary, this work introduces a method for constructing quantum circuits by manipulating magnon wave packets on spin chains through controlled alignment of their magnetization directions. By setting the chains in parallel or anti-parallel configurations, the system can function as a magnon squeezer, which generates magnon entanglement, or as a magnon beam splitter, respectively. Using the quantum Langevin equation, simulations of entanglement generation, verification, and quantum teleportation protocols were performed on double- and triple-chain quantum circuits. These results highlight the promise of magnon-based approaches for advancing quantum information technologies.

\emph{Acknowledgements. } 
This work was supported by 
the National Key Research and Development Program of China (Grant No. 2022YFA1403300), 
the National Natural Science Foundation of China (Grants No. 12474110), 
The Quantum Science and Technology - National Science and Technology Major Project
(Grant No.2024ZD0300103),
and Shanghai Municipal Science and Technology Major Project (Grant No.2019SHZDZX01).

\bibliographystyle{apsrev4-2}
\bibliography{Magnon_teleportation}

\newpage 

\onecolumngrid

\appendix 

\section{A. Drift Matrix}
\label{Appendix A}

Here we show how to obtain the drift matrix of the Gaussian dynamics in the double chain system. We start from the Heisenberg equation of motion
\begin{equation}
    \dv{t}\hat{M}_{\alpha i s} = \ii [\hat H, \hat{M}_{\alpha i s}]
    = \ii D_{\alpha i s}^{\alpha'i's'} \hat{M}_{\alpha' i' s'},
\end{equation}
where 
\begin{subequations}
\begin{align}
    \hat H &= \hH_0^{\text{eff}} + \hH_c^{\text{P}} + \hH_c^{\text{AP}} \\
    \hH_0^{\text{eff}} &= \sum_{\alpha, i} K \hat{M}_{\alpha i +}\hat{M}_{\alpha i -} 
    + J \qty(\hat{M}_{\alpha i +}\hat{M}_{\alpha (i+1) -} + \hat{M}_{\alpha i -}\hat{M}_{\alpha (i+1) +}),  \\
    \hH_c^{\text{P}} &= J'\sum_{i\in\mathcal C_{\text{P}}} \hat{M}_{A i +}\hat{M}_{B i -} +  \hat{M}_{A i -}\hat{M}_{B i +}
    \qand
    \hH_c^{\text{AP}} = J'\sum_{i\in\mathcal C_{\text{AP}}} \hat{M}_{A i +}\hat{M}_{B i +} + \hat{M}_{A i -}\hat{M}_{B i -}  
\end{align}
\end{subequations}
By applying the canonical commutation relation 
\[
\comm{\hat{M}_{\alpha i s}}{\hat{M}_{\alpha' i' s'}}
= s\Delta_{\alpha i \overline{s}}^{\alpha' i' s'}
= s\begin{cases}
1 & \alpha'i's' = \alpha i \bar{s} \\
0 & \text{otherwise}
\end{cases},
\]
we obtain the expression of the drift matrix
\begin{subequations}
    \begin{align}
    D&= {^{0}D} + {^{\text{P}}D} + {^{\text{AP}}D}, \\
        {^0D}_{\alpha i s}^{\alpha'i's'} 
        &= \dv{[\hat H_0^{\text{eff}},\hat{M}_{\alpha i s}]}{\hat{M}_{\alpha' i' s'}}  
        = s K \Delta_{\alpha i s}^{\alpha'i' s'} 
        + s J \qty(\Delta_{\alpha (i+1) s}^{\alpha' i' s'} + \Delta_{\alpha (i-1) s}^{\alpha' i' s'} ), \\
        {^{\text{P}}D}_{\alpha i s}^{\alpha'i's'} 
        &= \dv{[\hat H_c^{\text{P}},\hat{M}_{\alpha i s}]}{\hat M_{\alpha' i' s'}} 
        = s J' \Delta_{\overline{\alpha} i s}^{\alpha' i' s'}
        \qand
        {^{\text{AP}}D}_{\alpha i s}^{\alpha'i's'} 
        = \dv{[\hat H_c^{\text{AP}},\hat M_{\alpha i s}]}{\hat M_{\alpha' i' s'}} 
        = s J' \Delta_{\overline{\alpha} i \overline{s}}^{\alpha' i' s'}.
    \end{align}
\end{subequations}

For the case with inhomogeneous magnetic texture such as a domain wall, 
the effective Hamiltonian is adjusted by the spatially varying magnetic configuration parametrized by the local equilibrium magnetic moment angle $\theta_\alpha^i$, 
\[
\hH^{\text{eff}}_{\theta} 
= \sum_{\alpha,i} -\frac12K \qty[\hat z \cdot R(\theta_{\alpha}^{i})\mb_{\alpha}^{i}] 
- J \mb_{\alpha}^{i} \cdot R(\Delta \theta_{\alpha}^i) \mb_{\alpha}^{i+1}. 
\] 
Here $R(\theta)$ is the rotation matrix in the $xz$ plane. In terms of $M_{\alpha i s}$, the magnon operator is expressed as 
\[
\mb_{\alpha}^{i} = \qty(\frac1{\sqrt2}(M_{\alpha i -} + M_{\alpha i +} ), \frac1{\ii \sqrt2}(M_{\alpha i -} - M_{\alpha i +} ), 1- M_{\alpha i +}M_{\alpha i -} ).
\] 
We assume the domain wall is on chain-B, then $\theta_{A}^{i} = 0$ and $\theta_{B}^{i}$ is denoted as $\theta_i$. After omitting the constant terms and anharmonic terms, we obtain the addition Hamiltonian associated with the domain wall as
\begin{align}
    \hH^{\text{DW}}  =& \hH^{\text{eff}}_{\theta} - \hH^{\text{eff}}_{0} \nonumber \\
    =& \sum_{i} -\frac12 K \sin^2\theta_i \qty(3\hat M_{B i +} \hat M_{B i -} + \hat M_{B i +} \hat M_{B i +} + \hat M_{B i -} \hat M_{B i -}) \nonumber\\
    & + J \sin^2\frac{\Delta\theta_i}{2}\qty[ \hat M_{B i +} \hat M_{B i -} 
    - \qty( \hat M_{B i +} \hat M_{B (i+1) -} + \hat M_{B i -} \hat M_{B (i+1) +} )
    + \qty( \hat M_{B i +} \hat M_{B (i+1) +} + \hat M_{B i -} \hat M_{B (i+1) -} )
    ]
\end{align}
For the case of a Walker type domain wall stabilized by uniaxial anisotropy $K$ along $\hbz$ and Heisenberg exchange coupling $J$, we have $\frac K2\sin2\theta_i = J\qty(\sin\Delta\theta_{i+1}-\sin\Delta\theta_{i})$.
$\hH^{\text{DW}}$ gives rise to an additional term in the drift matrix 
\begin{align}
    {^{\text{DW}}D}_{\alpha is}^{\alpha'i's'} 
    = \dv{[\hat H^{\text{DW}},\hat M_{\alpha i s}]}{\hat M_{\alpha' i' s'}} 
    &=  -\frac12 s K \sin^2\theta \qty( 3 \Delta_{\alpha i s}^{\alpha'i' s'} + 2 \Delta_{\alpha i \overline{s} }^{\alpha'i' s'} ) \nonumber \\
    & + s J \sin^2\frac{\Delta\theta_i}{2}\qty[ \Delta_{\alpha i s}^{\alpha'i' s'} 
    +  \qty(\Delta_{\alpha (i+1) \overline{s}}^{\alpha' i' s'} + \Delta_{\alpha (i-1) \overline{s}}^{\alpha' i' s'} )
    +  \qty(\Delta_{\alpha (i+1) s}^{\alpha' i' s'} + \Delta_{\alpha (i-1) s}^{\alpha' i' s'} )
    ]. 
\end{align}
For sites outside the domain wall region, $\theta_i = 0 $, makes the contribution of ${^{\text{DW}}D} = 0$ in these regions.

\section{B. Finite Temperature and Dissipation}

For the finite temperature and the finite dissipation, a dissipative matrix and a thermal noise should be included in the Langevin equation: 
\begin{equation}
    \dv{t} M(t) = (D-\Gamma)  M_(t) 
    +\sqrt{2 n(T)+1} h(t),
\end{equation}
where $n(T) = [\exp(\hbar K / k_B T)-1]^{-1}$ is the thermal magnon number, $\Gamma_{\alpha is}^{\alpha'i's'}=\gamma \Delta_{\alpha i}^{\alpha'i's'}$ is the damping matrix with uniform damping rate $\gamma$ for all sites. 
The equation of motion for the covariant matrix should be revised as
\begin{equation}
    \label{eqn:dVdtT}
      \dv{t}V(t) = \tilde D V(t) + V(t)\tilde D^T + [2n(T) + 1]\Gamma
      \qwith \tilde D = D - \Gamma.
\end{equation}

\begin{figure}[ht]
    \includegraphics[width=0.8\columnwidth]{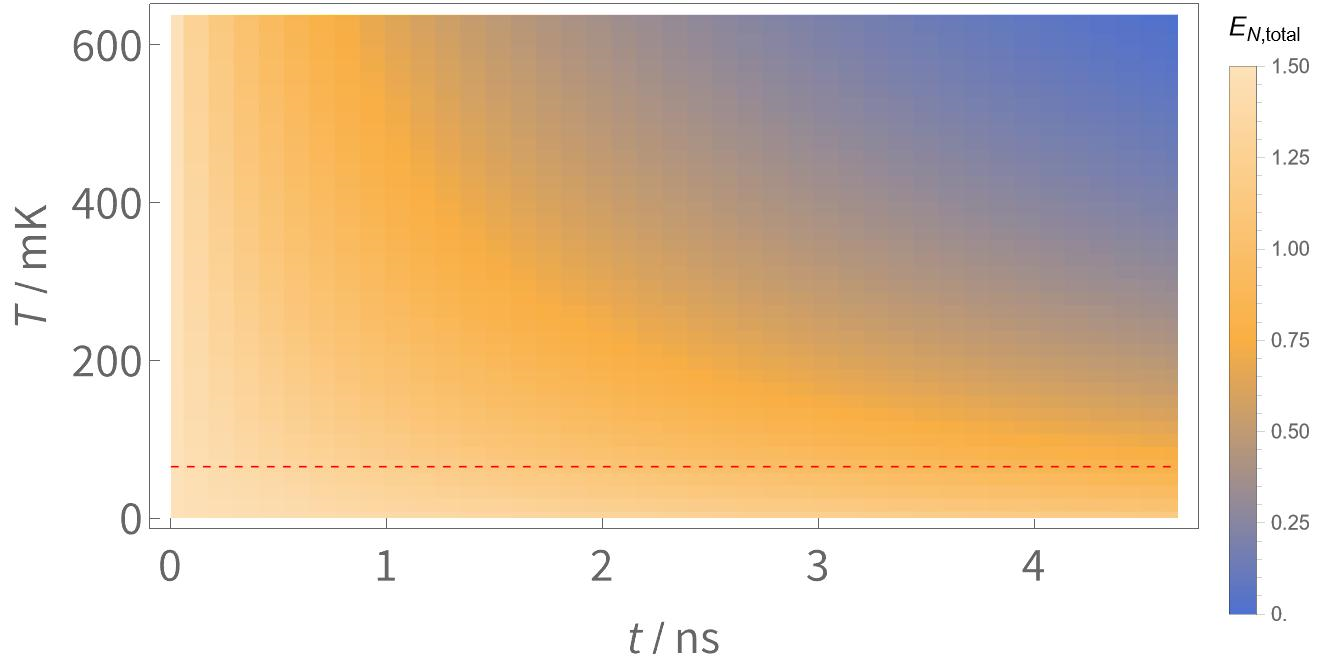}
    \caption{The total inter-chain entanglement of a double chain model in \Figure{fig:doublechain}, $E_{\text{N,total}}=E_{\text{N}}[V]$, is calculated as a function of time at different temperatures. The damping rate is fixed at $\gamma = 10^{-4}$. The P coupling region is switched off here. The red dashed line is $\SI{65}{mK}$, correlated to a single magnon excitation in YIG.}
    \label{fig:temperature_damping}
\end{figure}

Both of the finite temperature and the dissipation lead to the dephasing of the quantum state, and contribute to the reduction of the entanglement. 
\Figure{fig:temperature_damping} shows the entanglement across the double chain, calculated using the covariant matrix obtained from \Eq{eqn:dVdtT}, as function of time and temperature, showing the diminishing of overall entanglement with increasing temperature and elapse of time. The damping rate is fixed at $\gamma = 10^{-4}$, corresponding to common Gilbert damping constant for YIG.

\section{C. Teleportation Fidelity}

To exam how the teleportation fidelity depends on the coupling phases $\phi_{\text{AP}} = \omega_{\text{AP}}\tau_{\text{AP}}$ and $\phi_{P} = J'\tau_{\text{P}}$ in the P and AP region, we first solve the QLE \Eq{eqn:dVdt} to obtain the first and second-order moments, $M(t)$ and $V(t)$, for specific coupling time $\tau_{\text{AP}}$ in the AP region and the length $l_{\text{P}}$ of the P region in \Figure{fig:tel}(a). 
The initial state of the triple-chain for solving \Eq{eqn:dVdt} is a direct product state of a coherent state wave packet on chain-C (with Wigner distribution $W_C(M_C,0)$) and the vacuum state on chain-A and -B (with Wigner distribution $W_{AB}(M_A,M_B,0)$):
\[ W(\mu,0)=W_{\text{C}}(M_C,0)W_{\text{AB}}(\mu_A,\mu_B,0). \]

Once we have the first and second-order moments $M(t)$ and $V(t)$ for all times, we may construct the Wigner function for the triple-chain as:
\begin{equation}
    \label{eqn:WMt}
    W(\mu,t) = \frac{1}{\sqrt{\det V(t)}} \exp\qty{-\frac12 [\mu-\ev*{M(t)}]^T V(t)^{-1} [\mu - \ev*{M(t)}]}, 
\end{equation}
Here vector $\mu$ is the variable of the Wigner function with index $\alpha i s$, $\ev*{M(t)}$ is the first order moment of the random variable $M(t)$, and $V(t)$ is the covariant matrix. All indices are omitted for simplicity. 

The homodyne detections on chain-A and -C (along $M_{A i +}$ and $M_{C i -}$ for example) lead to the collapse of the states on both chain-A and -C collapse, and yield results $\nu_{A+}$ and $\nu_{C-}$, respectively. The Wigner function for the teleported state on chain-B is obtained from \Eq{eqn:WMt} by convolution \cite{Quantum_information_with_continuous_variables_Review_2005}
\begin{equation}
    W_{\text{tel}}(\mu_B) = \iint \dd{\mu_A} \dd {\mu_C} W(M,t) \nn
    \delta(\mu_{A i +}-\nu_{A+}) \delta(\mu_{C i -}-\nu_{C-}),
\end{equation}
where $\dd \mu_A\dd \mu_C = \prod_{i=1}^N\prod_{s=\pm1}\dd \mu_{A i s}\dd \mu_{C i s}$.

The teleportation fidelity is obtained by the inner product between the teleported state and the initial state represented by Wigner function $W_C(M_C,0)$ injected into chain-C
\[ f(\phi_P,\phi_{AP})= \int \dd \mu ~ W_\text{C}(\mu_C,0)W_{\text{tel}}(\mu_B)\delta(\mu_B-\mu_C). \] 
The fidelity shown in \Figure{fig:tel}(e) is plotted using this expression calculated for scenarios with different phase accumulation in P and AP region.

\end{document}